\documentclass{emulateapj}

\usepackage{amsmath,amstext}
\usepackage{natbib}
\usepackage{amsfonts}
\usepackage[colorlinks=true,citecolor=blue]{hyperref}
\usepackage{graphicx}
\usepackage{epstopdf}
\usepackage[export]{adjustbox}
\def\na{New A}                % New Astronomy

\shorttitle{Kozai-Lidov Mechanism and Black Hole Mergers}
\shortauthors{VanLandingham et al.}

\begin{document}

\title{The Role of the Kozai-Lidov Mechanism in Black Hole Binary Mergers in Galactic Centers }
\author{John H. VanLandingham}
\author{M. Coleman Miller}
\author{Douglas P. Hamilton}
\author{Derek C. Richardson}
\affil{Department of Astronomy, University of Maryland, College Park, MD 20742-2421, USA}

\begin{abstract}
In order to understand the rate of merger of stellar-mass black hole binaries (BHBs) by gravitational wave (GW) emission it is important to determine the major pathways to merger. We use numerical simulations to explore the evolution of BHBs inside the radius of influence of supermassive black holes (SMBHs) in galactic centers. In this region the evolution of binaries is dominated by perturbations from the central SMBH. In particular, as first pointed out by Antonini and Perets, the Kozai-Lidov (KL) mechanism trades relative inclination of the BHB to the SMBH for eccentricity of the BHB, and for some orientations can bring the BHB to an eccentricity near unity. At very high eccentricities, GW emission from the BHB can become efficient, causing the members of the BHB to coalesce. We use a novel combination of two \emph{N}-body codes to follow this evolution. We are forced to simulate small systems to follow the behavior accurately. We have completed 400 simulations that range from $\sim$ 300 stars around a $10^{3}$ M$_{\odot}$ black hole to $\sim$ 4500 stars around a $10^{4}$ M$_{\odot}$ black hole. These simulations are the first to follow the internal orbit of a binary near a SMBH while also following the changes to its external orbit self-consistently. We find that this mechanism could produce mergers at a maximum rate per volume of $\sim 100$~Gpc$^{-3}$~yr$^{-1}$ or considerably less if the inclination oscillations of the binary remain constant as the BHB inclination to the SMBH changes, or if the binary black hole fraction is small. 
\end{abstract}


\maketitle

\section{Introduction}

The recent detection of gravitational waves (GWs) from the merger of a black hole binary (BHB) by Advanced LIGO \citep{PhysRevLett.116.061102} has begun a new phase of investigation in GW astronomy. Instead of asking whether detectors such as Advanced LIGO \citep{2015CQGra..32g4001L} and Advanced Virgo \citep{2015CQGra..32b4001A} will detect GWs, or whether BHBs exist, we turn to determining how common mergers of BHBs are and what pathways to merger they follow. Currently, predictions for the merger rate of BHBs are very uncertain. Until this detection, there had been no observations of merging black holes to inform the predictions. Instead, various population synthesis models \citep{2003MNRAS.342.1169V,2007PhR...442...75K,2014LRR....17....3P,2015arXiv151004615B} have been used to predict the merger rate. The predicted rates per volume range from $0.1$ Ð- $300$ Gpc$^{-3}$ yr$^{-1}$ \citep{2010CQGra..27q3001A}. The recent gravitational wave detection has constrained this rate to $2$ -- $400$ Gpc$^{-3}$ yr$^{-1}$, which eliminates only the lowest values \citep{PhysRevLett.116.061102}.

\indent Population synthesis models typically rely on stellar evolution codes to predict the number and distribution of black hole binaries. Dynamical interactions with other stars or black holes may increase the predicted merger rates. These interactions have recently begun to be studied, particularly in dense stellar environments such as globular clusters and galactic centers \citep{2000ApJ...528L..17P,2002ApJ...576..894M,2006ApJ...637..937O,2008ApJ...676.1162S,2009ApJ...692..917M,2012ApJ...757...27A,2015ApJ...800....9M,2015PhRvL.115e1101R,2016arXiv160202444R}. The environment of the nuclear region of galaxies is of particular interest as secular processes due to the presence of the central supermassive black hole (SMBH) can become important. 

\indent In galactic nuclei, within the region where the SMBH dominates the gravitational potential (the radius of influence), a BHB would effectively form a hierarchical triple system with the SMBH, where the BHB would to first order be simply orbiting the SMBH. Under these conditions, as long as the BHB has a high relative inclination to its orbit around the SMBH, it will evolve over many orbits trading eccentricity for inclination in a periodic fashion \citep{2010ApJ...713...90A,2012ApJ...757...27A}. This mechanism was first explored by \citet{1962AJ.....67..591K} and \citet{1962P&SS....9..719L} with a focus on asteroids being perturbed by Jupiter, and is referred to as the Kozai-Lidov (KL) mechanism. The KL mechanism has since been further developed to higher order and less restricted cases \citep[e.g.,][]{1976CeMec..13..471L,1997AJ....113.1915I,2000ApJ...535..385F,2002ApJ...576..894M,2002ApJ...578..775B,2011ApJ...742...94L}. 

\indent The KL mechanism, when applied to BHBs in galactic nuclei, can lead to binaries with very high eccentricities, and therefore to a merger timescale that is very short compared to a circular binary. As an example, consider a system consisting of two $10$ M$_{\odot}$ black holes orbiting a SMBH. From \citet{1964PhRv..136.1224P} if the binary has a separation of 1 AU it would take $~10^{14}$ years to merge due to gravitational wave emission for a circular binary, but only four thousand years to merge for a binary with $e = 0.9995$. Reaching this ÒextremeÓ eccentricity is possible due to the KL mechanism if the initial mutual inclination between the BHB and SMBH is approximately 88--92 degrees \citep{1962AJ.....67..591K}.

\indent Determining the rate of BHB mergers due to the KL mechanism in a galactic nucleus is complicated by the various other processes affecting the orbit of the BHB in such a dense stellar environment. These include such effects as two-body relaxation \citep{1987degc.book.....S}, scalar and vector resonant relaxation \citep{1996NewA....1..149R,2006ApJ...645.1152H,2015MNRAS.448.3265K}, mass segregation \citep{1977ApJ...216..883B,2006ApJ...645L.133H,2014ApJ...794..106A}, binary evaporation \citep{1987gady.book.....B}, general relativistic precession, and close approaches \citep{1977ComAp...7...43H}. Several of these processes would tend to suppress KL cycles, whereas others offer the tantalizing possibility of increasing the chance that a BHB will end up in a favorable orientation. 

\indent To date, studies of binaries undergoing KL oscillations in galactic centers \citep[e.g.,][]{2010ApJ...713...90A,2012ApJ...757...27A,2015ApJ...799..118P,2016arXiv160302709S} have assumed that the mutual inclination of the binary and SMBH is fixed at some initial value. However, as the binary orbits the SMBH the inclination of the center of mass (COM) of the binary relative to the SMBH will be altered due to the asymmetric potential of stars the binary orbits within, as well as by close approaches to the binary. If this change in the inclination of the COM orbit of the binary can lead to even minor changes in the total inclination between the binary and SMBH, this will result in many more binaries reaching the critical inclination that allows them to merge in a relatively short time. 

\indent In order to follow all of these processes we turn to simulations of galactic nuclei using numerical $N$-body gravity codes. Many advances have been made in $N$-body simulations of dense stellar systems. Highly parallelized codes \citep{2001PhDT........21S,2012MNRAS.424..545N} and extremely fast codes using GPUÕs \citep{2007NewA...12..641P,2009NewA...14..630G,2015MNRAS.450.4070W} allow for simulations with great precision and large numbers of particles \citep{2012MNRAS.425.2872H,2013MNRAS.430L..30S,2014MNRAS.445.3435H}. However, difficulties remain, particularly with highly eccentric orbits near very massive objects. These situations require very high precision because of the extreme mass ratios involved, as well as the very large forces, which require very small time steps to integrate correctly. Simulations of this kind are possible but have often been limited to small numbers of particles \citep{2008AJ....135.2398M,2011PhRvD..84d4024M}. This difficult situation is precisely the regime that we would like to explore. The additional requirement of following the internal orbit of a close binary in this regime leads us to use novel methods of simulation and restrict ourselves to systems that have SMBHs that are a few orders of magnitude smaller than is realistic. 

\indent Here we describe full $N$-body simulations of BHB mergers in dense stellar regions around a massive black hole. Our simulations assume that the changes in the mutual inclination between the BHB and SMBH follow the inclination of the COM orbit of the BHB. Recent work by \citet{2015MNRAS.449.4221H} as well as our own tests suggest that under certain circumstances this should not be the case. Our results should therefore be considered an upper limit of the case in which mutual inclination changes are damped or eliminated. Our simulations are evidence of the potential importance of the contribution to the BHB merger signal for Advanced LIGO by binaries that merge due to the KL mechanism. Additionally, as we discuss in Section 5, these simulations provide intriguing insight into observations of the possible overabundance of low mass X-ray binaries very close to the galactic center \citep{2005ApJ...622L.113M,2015ApJ...799..118P,2015Natur.520..646P}. 

\indent In Section 2 we set up our problem in more detail, including relevant timescales for our simulations. Section 3 discusses the problems associated with simulating dense, highly eccentric groups of stars on these timescales. We also describe the simulation method we use to solve those problems. In Section 4 we present our results, along with a discussion of the implied detection rate of mergers. Finally, in Section 5 we address various physical processes that may increase or decrease the binary merger rate and that may be incompletely simulated, and we present our conclusions in Section 6.

\section{Timescales}

Our focus in this paper is on stellar mass black hole binaries (BHBs) in galactic centers, inside the radius of influence of the supermassive black hole (SMBH). In this section, we examine the various processes that may affect this binary, with a particular view towards the eventual merger of these binaries and the resulting gravitational wave emission.

\indent The gravitational wave merger timescale is given by \citet{1964PhRv..136.1224P} as: 
\begin{multline}\label{eq:grav}
T_{\rm GW} \simeq \frac{3}{85}\left(\frac{c^{5}a_{0}^{4}}{G^{3}m_{1}m_{2}M_{\rm b}}\right)(1-e_{0}^2)^{7/2} \\ 
\simeq 1.2 \times 10^{14} \mbox{yr} \left(\frac{M_{\rm b}}{20 \mbox{M}_{\odot}}\right)^{-3}\left(\frac{a_{0}}{1 \mbox{AU}}\right)^{4}(1-e_{0}^2)^{7/2}\mbox{  .}
\end{multline}

\noindent Here $a_{0}$ is the semi-major axis of the binary, $e_{0}$ is the eccentricity, $M_{\rm b}$ is the binary mass, and $m_{1}$ and $m_{2}$ are the components. The second line assumes an equal mass binary. A circular binary consisting of two 10 M$_{\odot}$ black holes with a semi-major axis of 1 AU would require much more than the age of the universe to merge by gravitational wave emission alone. However, there is a strong dependence on the eccentricity. A binary with the same semi-major axis and component masses but with an eccentricity of 0.9995 would merge in a mere 3800 years. It is therefore clear that a process that can significantly increase the eccentricity of binary systems could dramatically increase the rate of mergers.

\begin{figure}[htb]
\centering
\includegraphics[height = 3in, width = 3.5in]{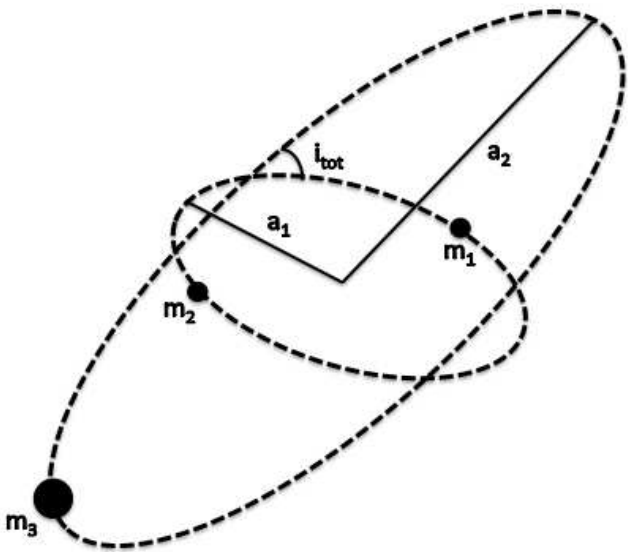}
\caption{\label{fig:coords} \small{Coordinate system for the general Kozai problem. In our case $m_{1}$ and $m_{2}$ are the components of the BHB while $m_{3}$ is the SMBH.  The mutual inclination is $i_{tot}$, $a_1$ is the interior semi-major axis, and $a_2$ is the superorbit semi-major axis. Adapted from \citet{2013MNRAS.431.2155N}}}
\end{figure}

\indent One important process that can bring binaries to extremely high eccentricities is the Kozai-Lidov (KL) mechanism. The KL mechanism is an application of the three-body problem to the special case of a hierarchical triple system. In a hierarchical triple system, an inner binary with small semi-major axis is orbited by a third body at a much larger semi-major axis. More directly instructive for our purposes, a hierarchical triple system is one in which the outer object dominates the angular momentum of the system. This is exactly the situation for a BHB near a SMBH. The components of the BHB, $m_{1}$ and $m_{2}$, form the inner binary, which from its frame of reference is orbited at a large distance by the SMBH, $m_{3}$ (see Figure \ref{fig:coords} for coordinate system). In the rest of this paper we will treat the SMBH as stationary with the BHB orbiting it, which is equivalent to this formalization but a more natural frame of reference. 

\graphicspath{{./Figures/}}

\indent This particular case of the hierarchical three-body problem can be solved analytically by expanding the Hamiltonian in a power series in $a_{1}/a_{2}$ where $a_{1}$ is the semi-major axis of the inner binary or interior orbit, and $a_{2}$ is the semi-major axis of the outer binary, which we call the superorbit. The main result of these analyses is that $i_{tot}$, the mutual inclination of the inner and outer binary, will oscillate over many orbits of the inner binary along with $e_{1}$, the eccentricity of the inner binary. That is, a system with large $i_{tot}$ and small $e_{1}$ will evolve to a system with small $i_{tot}$ and large $e_{1}$.

\indent The first analyses of this mechanism \citep{1962P&SS....9..719L,1962AJ.....67..591K} considered the effect of the perturbation of the orbit of a highly inclined asteroid by the gravity of Jupiter. These analyses solve the equations of motion to quadrupole order and assume a circular outer orbit ($e_{2} = 0$) that dominates the angular momentum of the three-body system (i.e., $m_{2} = 0$). In this regime, a simple relation can be found between the maximum eccentricity of the inner binary ($e_{1,max}$) and the initial mutual inclination ($i_{\rm{tot},0}$) provided that $39^{\circ} \lesssim i_{\rm{tot},0} \lesssim 141^{\circ}$. This is given by \citep{1962AJ.....67..591K}:
\begin{equation}
e_{1,\rm max} \approx \left(1-\frac{5}{3}\cos^2i_{\rm{tot},0}\right)^{1/2}.
\end{equation}
A system with initial mutual inclination of $90^{\circ}$ would allow the inner binary to reach an eccentricity of unity. In this limit these oscillations can be simply understood as the conservation of the component of the angular momentum of the inner orbit along the direction of the total angular momentum of the system.

\indent Later analyses of this effect remove the restrictions of a circular outer binary and test mass particle \citep{1976CeMec..13..471L}, take the approximation to octupole order \citep{2000ApJ...535..385F}, and add general relativistic effects \citep{2002ApJ...578..775B,2002ApJ...576..894M}. All of these analyses find the same general result, that at high enough initial inclinations, the eccentricity and inclination of the inner orbit will oscillate. The exact details of the maximum eccentricity and the initial inclination needed may vary, and at the octupole level the inclination can even change sign \citep[e.g.,][]{1997AJ....113.1915I,2011ApJ...742...94L}. We take care of all of these details automatically by doing direct three-body numerical integrations.

\indent In order for the KL mechanism to become important for a BHB, several conditions must be met. First, the BHB must be in a hierarchical triple system with some other object. Second, the BHB must reach a critical inclination such that KL oscillations are strong enough to reach high eccentricity. Third, the BHB must remain in the correct orientation long enough that the KL oscillations can allow the BHB to merge before other processes suppress them. The first condition is met by considering BHBs that orbit within the radius of influence of a SMBH. The radius of influence $R_{\rm infl}$ is defined as the radius at which the velocity dispersion of the bulge of the galaxy $\sigma$ is equal to the speed expected of a Keplerian orbit around the SMBH,

\begin{equation}
R_{\rm infl} = \frac{GM_{\rm SMBH}}{\sigma^{2}} \approx 1\mbox{pc} \left(\frac{M_{\rm SMBH}}{10^{6}M_{\odot}}\right)^{1/2}.
\end{equation}

\noindent Here the second equality is found by using the $M$ -- $\sigma$ relation, which is an empirical relation between the mass $M_{\rm SMBH}$ of a SMBH and the velocity dispersion $\sigma$ of the galaxy bulge. This relation has been estimated to be  $M_{\rm SMBH}\propto\sigma^\alpha$, with $\alpha\sim 4-5$ \citep[e.g.,][]{2000ApJ...539L...9F,2002ApJ...574..740T,2009ApJ...698..198G}. Here we simply use $\alpha = 4$. Alternatively, $R_{\rm infl}$ can be defined as the radius inside of which the mass in stars is equal to the mass of the SMBH. Empirical relations using this definition have been found \citep{2009ApJ...699.1690M,2016MNRAS.455..859S} and find a similar relation, though with a slightly different slope. The binaries under consideration here are well inside $R_{\rm infl}$.

\indent The conditions that the BHB reach a critical orientation as well as remain in that orientation long enough for merger are much more difficult to predict. \citet{2012ApJ...757...27A} examine many of the timescales relevant for this analysis. In Table \ref{tab:time} we provide the most relevant of these timescales, adapted for our particular circumstances. In their paper, Antonini \& Perets conduct an analysis of the appropriate timescales as well as a series of three-body numerical integrations of BHBs near a SMBH and conclude that the KL mechanism is likely to increase the number of BH-BH mergers in galactic nuclei. 

\begin{table*}[htp]
\renewcommand{\arraystretch}{5}
\begin{center}
\caption{\label{tab:time}}
\begin{tabular}{|c||c|}
\hline
Timescale & Note\\
\hline \hline
\(\displaystyle T_{\rm KL} = 1.4 \times 10^{6} \mbox{yr} \left(\frac{M_{\rm SMBH}}{10^{6} \mbox{M}_{\odot}}\right)^{-1}\left(\frac{M_{\rm b}}{20 \mbox{M}_{\odot}}\right)^{1/2}\left(\frac{a_{1}}{1 \mbox{AU}}\right)^{-3/2}\left(\frac{a_{2}}{0.1 \mbox{pc}}\right)^{3}(1-e_{2}^2)^{3/2}\) & 1\\
\(\displaystyle T_{\rm ER} = 4.4 \times 10^{7} \mbox{yr} \left(\frac{M_{\rm SMBH}}{10^{6} \mbox{M}_{\odot}}\right)\left(\frac{M_{\rm b}}{20 \mbox{M}_{\odot}}\right)^{-1}\left(\frac{a_{2}}{0.1 \mbox{pc}}\right)^{1/2}\) & 2\\
\(\displaystyle T_{\rm RR} = 2.3 \times 10^{7} \mbox{yr} \left(\frac{M_{\rm SMBH}}{10^{6} \mbox{M}_{\odot}}\right)^{1/2}\left(\frac{M_{\rm b}}{20 \mbox{M}_{\odot}}\right)^{-1}\left(\frac{a_{2}}{0.1 \mbox{pc}}\right)^{3/2}\) & 3\\
\(\displaystyle T_{\rm VRR} = 4.6 \times 10^{4} \mbox{yr} \left(\frac{M_{\rm SMBH}}{10^{6} \mbox{M}_{\odot}}\right)^{1/4}\left(\frac{M_{\rm b}}{20 \mbox{M}_{\odot}}\right)^{-1}\left(\frac{a_{2}}{0.1 \mbox{pc}}\right)\) & 4\\
\(\displaystyle T_{\rm GR} = 6.0 \times 10^{4} \mbox{yr} \left(\frac{M_{\rm b}}{20 \mbox{M}_{\odot}}\right)^{-3/2}\left(\frac{a_{1}}{1 \mbox{AU}}\right)^{5/2}(1-e_{1}^2)\) & 5\\
\hline
\end{tabular}
\end{center}
\tablecomments{\\
\small{1. Kozai-Lidov timescale \citep{1997AJ....113.1915I}. The timescale over which Kozai-Lidov oscillations occur. Here $a_{2} = 0.1R_{\rm infl}$ is the superorbit semi-major axis, $e_{2}$ is the superorbit eccentricity, and $a_{1}$ is the interior orbit semi-major axis.\\
2. Energy Relaxation timescale \citep{1987degc.book.....S}. The timescale over which objects can significantly alter their orbital energy. This is the same process as dynamical friction. We use a Keplerian velocity dispersion since our region of interest is inside $R_{\rm infl}$. The $M$--$\sigma$ relation and a number density of stars $n \propto r^{-\alpha}$ with $\alpha = 2$ sets the number of stars within the volume. We choose $\alpha$ primarily to speed up the simulations, though a value of $\alpha \approx 2$ is physically motivated \citep[e.g.,][]{2003ApJ...594..812G}. The average stellar mass is set to 1 $\mbox{M}_{\odot}$.\\
3. Resonant Relaxation timescale \citep{1996NewA....1..149R}. The timescale over which objects can significantly alter their angular momentum.\\
4. Vector Resonant Relaxation timescale \citep{2006ApJ...645.1152H}. VRR is due to the averaged mass distribution over many orbits of an individual star, which exerts a torque that can alter the plane of the BHB orbit without affecting its energy or the magnitude of its angular momentum. Here once again the $M$--$\sigma$ relation and stellar number density are used to set the number of stars in the volume.\\
5. General Relativistic Precession timescale. If this precession is too fast, it can suppress the KL cycles. Previous work \citep{2002ApJ...578..775B,1997ApJS..112..423H} has found the precise condition at which this happens. In our simulations this condition is rarely met.}}
\end{table*}

\subsection{Mutual Inclination} 

The single most important factor for whether a BHB will merge while in the high eccentricity phase of a KL oscillation is its mutual inclination with the SMBH, $i_{\rm{tot}}$. If $i_{\rm{tot}}$ is close enough to $90^{\circ}$, the BHB will reach a high enough eccentricity that it can merge during a single KL oscillation (See Equations 1 and 2). It is therefore critical to know the evolution of $i_{\rm{tot}}$ and any processes which might affect it. Research done prior to this on the importance of the KL mechanism to binaries has sometimes mentioned the possibility that $i_{\rm{tot}}$ may not be fixed \citep[e.g.,][]{2009ApJ...699L..17P,2012ApJ...757...27A,2016ApJ...816...65A,2016arXiv160302709S}, but simulations including this effect have not been completed. If indeed there are processes which could change $i_{\rm{tot}}$, this would expand the number of binaries which could reach critical inclination dramatically, possibly allowing many more BHBs to merge.

\indent Any process that changes the orbital plane of the BHB around the SMBH has the potential to alter $i_{\rm{tot}}$. This includes vector resonant relaxation (VRR, see Table \ref{tab:time}) as well as precession due to the overall aspherical mass distribution of the galactic nucleus and kicks due to close approaches of stars. However, a recent paper by \citet{2015MNRAS.449.4221H} casts doubt on the efficacy of these processes ability to alter $i_{\rm{tot}}$. \citet{2015MNRAS.449.4221H} examine a series of 4-body systems that consist of a hierarchical triple system orbited at large distance by a fourth body. This fourth body, initially at high inclination relative to the inner triple system, is in effect causing a second KL oscillation on a different timescale from the inner triple. They find that as long as the timescale of the KL oscillations from the fourth body is long compared to the inner KL timescale, the mutual inclination of the inner triple is essentially unaffected by the fourth object. 

\indent In analogy with the system as described in Figure \ref{fig:coords}, this fourth object would be akin to a process that would alter the inclination of the superorbit of the BHB. According to this result, as long as the timescale of this process is slow compared to $T_{\rm KL}$, $i_{\rm{tot}}$ would remain fixed. In the opposite case, when the inclination altering process under consideration is fast relative to $T_{\rm KL}$, $i_{\rm{tot}}$ should closely follow the inclination of the superorbit. 

\indent In order to test this result in a situation more closely resembling our own scenario we performed a small set of 3-body integrations. We use the $N$-body code \textsc{hnbody} which is described in more detail in section 3. All of these simulations consist of a central object $m_{1}$ with mass 1 M$_{\odot}$ orbited at a distance of 1 AU by an object $m_{2}$ with mass $10^{-9}$ M$_{\odot}$ and at a distance of 10 AU by an object $m_{3}$ with mass $10^{-3}$ M$_{\odot}$. $T_{\rm KL}$ for this system is approximately $10^{6}$ yrs and the simulations were run for $3\times10^{6}$ yrs. We apply a force on $m_{3}$ which causes the inclination of its orbit around $m_{1}$ relative to a reference plane to increase exponentially over the first $10^{6}$ yrs.

\begin{figure}[htb]
\centering
\includegraphics[height = 3in, width = 3.5in]{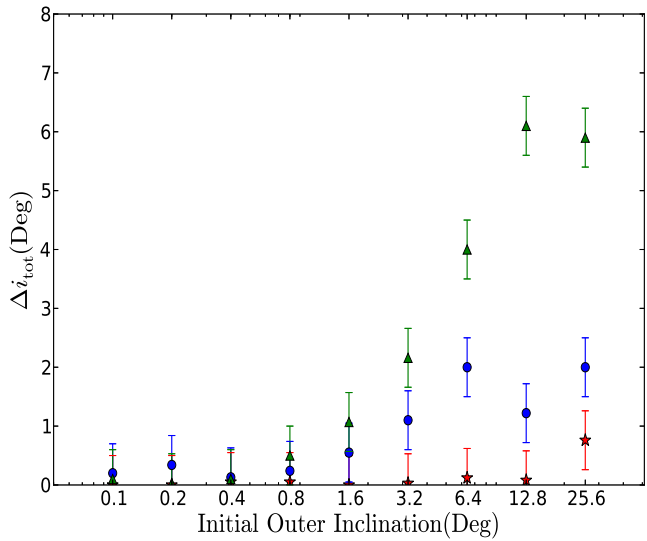}
\caption{\label{fig:mut} \small{Change in the mutual inclination of an inner and outer binary in a hierarchical triple as a function of the initial inclination of the outermost object, which increases its inclination by a factor of $e \approx 2.718$ over one KL cycle. The initial mutual inclination, $i_{\rm{tot}}$, is $55^{\circ}$ for red stars, $70^{\circ}$ for blue circles, and $85^{\circ}$ for green triangles. Error bars represent the variability of the inclinations. All values of $\Delta i_{\rm tot}$ within $0.5^{\circ}$ of $0^{\circ}$ are consistent with no change. Although $\Delta i_{\rm{tot}}$ is much smaller than the change in the inclination of the outermost object, it is not negligible, especially for large initial $i_{\rm{tot}}$.}}
\end{figure}

\indent Figure \ref{fig:mut} shows the results of these simple three body simulations. We performed three sets of simulations. The first set (red stars) are simulations where the initial mutual inclination, $i_{\rm{tot}}$, between the orbit of $m_{2}$ around $m_{1}$ and $m_{3}$ around $m_{1}$ is $55^{\circ}$. The second set (blue circles) have $i_{\rm{tot}} = 70^{\circ}$, while the third set (green triangles) have $i_{\rm{tot}} = 85^{\circ}$. The initial inclination of the outer object, $m_{3}$, ranges from $0.1^{\circ} - 25.6^{\circ}$, which means that over the first $10^{6}$ yrs the outer inclination changes by $\sim 0.3^{\circ} - 70^{\circ}$. This range of values corresponds to the range observed within one KL oscillation for our full $N$-body simulations. The values of $\Delta i_{\rm{tot}}$ are uncertain to within about $0.5^{\circ}$ due to the variability over time. All values of $\Delta i_{\rm tot}$ within $0.5^{\circ}$ of $0^{\circ}$ are consistent with no change.

\graphicspath{{./Figures/}}

\indent The simulations in Figure \ref{fig:mut} clearly show that $i_{\rm{tot}}$ does not follow the inclination of the outer object. However, while the change in $i_{\rm{tot}}$ is severely damped, it is not zero. Systems with initial $i_{\rm{tot}} = 85^{\circ}$ could even reach $90^{\circ}$. These preliminary results suggest that $i_{\rm{tot}}$ is not completely fixed, even with a somewhat slowly changing superorbit inclination. In the case of VRR or precession due to an aspherical mass distribution we would expect a smooth, relatively slow change in the superorbit inclination that might follow these results. However, kicks due to close approaches of stars would happen on a much faster timescale and might be more effective in changing the mutual inclination. In the following sections, we describe $N$-body simulations that allow us to self-consistently follow all of the important processes discussed in this section except for this damped mutual inclination change. Our results should therefore be considered an upper limit to the importance of inclination variance on the likelihood of KL oscillations leading to merger. 

\section{Numerical Simulations}

Direct $N$-body simulations have long been used to simulate star clusters, as they are very accurate and require very few assumptions. Over time, improving software and hardware has permitted simulations of ever greater numbers of particles. Although cosmological simulations have surpassed a billion particles, stellar simulations have progressed more slowly. In the last few years, the use of parallelized codes and codes that use GPU processors have allowed the direct simulation of star clusters with several hundred thousands of stars \citep[e.g.,][]{2013MNRAS.430L..30S,2014MNRAS.445.3435H} and even up to 1 million stars \citep{2015MNRAS.450.4070W,2016arXiv160104227R}. The discrepancy is due to the much larger number of dynamical times that are necessary to simulate these dense stellar systems. These are remarkable achievements, allowing us to model real star clusters accurately, although only for a few clusters so far \citep{2011MNRAS.411.1989Z,2014MNRAS.440.3172Z,2014MNRAS.445.3435H}. 

\indent Another type of system that could benefit greatly from $N$-body simulations, and the one of interest here, is galactic centers. Here, however, direct $N$-body simulations have been limited to several tens of stars \citep[e.g.,][]{2008AJ....135.2398M,2011PhRvD..84d4024M,2014MNRAS.437.1259B}. The reason for this discrepancy is the SMBH at the center of the cluster. Stars approaching close to the SMBH experience extreme forces and therefore require extremely high accuracy to integrate. This results in very time-consuming simulations. One alternative to using a small number of stars is to use a relatively small SMBH mass or relatively large particle mass \citep[e.g.,][]{2012ApJ...745...83A,2014ApJ...785..163V}. The underlying problem is the extremely small time steps required for integrating close approaches of particles of very disparate masses.

\subsection{Methodology}

Our goal is to follow the orbit of a BHB under the influence of the perturbations from a SMBH as well as the effects of the stars in a galactic center. We therefore must follow not only close approaches of stars to the SMBH but also the internal orbit of the BHB. In order to achieve this we have developed a hybrid approach. We use the $N$-body code \textsc{nbody6} \citep{2012MNRAS.424..545N} to follow the orbit of the center of mass (COM) of the BHB as well as the orbits of all the stars and the SMBH. We use another $N$-body code \textsc{hnbody} \citep{1999DDA....30.0302R} to follow the three-body motion of the individual members of the binary under the influence of the SMBH. The procedure is as follows:

\begin{enumerate}
\item Generate a set of initial conditions for the stars, the SMBH, and a single COM BHB particle.
\item Integrate the stars and COM BHB particle forward until the BHB is close enough to the SMBH to be tidally separated using \textsc{nbody6} for each set of initial conditions.
\item Compile the position and velocity of the COM BHB particle for the duration of each simulation.
\item Generate a set of initial conditions for the initial orbit of the individual members of the BHB.
\item Integrate just the three-body interaction of the BHB and SMBH using \textsc{hnbody} for the duration of one output step from \textsc{nbody6}.
\item Reposition the COM position and velocity of the BHB calculated by \textsc{hnbody} to the position and velocity of the next output step from \textsc{nbody6}.
\item Continue integrating using \textsc{hnbody} and repositioning at each output step until the end of the \textsc{nbody6} run is reached or the BHB merges.
\end{enumerate}

Splitting our simulations into these separate parts allows us to follow the internal orbit of the BHB without slowing the simulation of all of the other stars. Without this novel approach, these simulations would not currently be possible. However, this also means that the changes in the COM orbit are implemented as instantaneous changes every \textsc{nbody6} output step rather than as smooth progressions. This is a good approximation for most effects because the \textsc{nbody6} output step is very short compared to the superorbit period. Unfortunately, this also means that we do not correctly follow the mutual inclination evolution as described in section 2.1. Instead, we effectively treat the mutual inclination as exactly following the changes in the superorbit inclination. The mutual inclination therefore varies more drastically than is realistic.

\indent Our approach also requires that the internal orbit of the BHB not be meaningfully affected by the influence of the other stars. One way in which a binary may be affected is by binary ionization. If a binary is soft \citep{1977ComAp...7...43H}, i.e., the magnitude of the internal binding energy of a binary is less than the typical kinetic energy of a star that comes close to the binary, then over many encounters the binary will tend to separate. Additionally, a very close encounter of a star within the actual orbit of the binary may randomize the binary orbit parameters. We therefore checked all of our binaries to see how common this was in our simulations. We found that our binaries averaged less than one close encounter per relaxation time. Because our binaries are hard and close encounters are rare, we feel the influence of individual stars on the internal elements of the BHBs is adequately modeled.\\

\subsection{Codes}

\textsc{nbody6} is a direct $N$-body integration code that uses fourth-order Hermite integration with hierarchical time stepping and several regularization schemes. For more information see \citet{2012MNRAS.424..545N}. We use \textsc{nbody6} to follow the COM orbit of our BHBs in the potential of the SMBH and stars. Using \textsc{nbody6} to simulate a galactic center requires following orbits very close to a mass much larger than the rest of the particles, a task for which the code was not designed. We use Kustaanheimo \& Stiefel (KS) regularization to help integrate close encounters with the SMBH. Still, it was necessary to keep the size of our simulations modest, as well as make a few adjustments to the code. 

\indent Because of the extreme mass ratio between stars and the SMBH in our simulations, it was possible for the time step of stars passing close to the SMBH to reach a value smaller than the precision of double precision floating points in \textsc{nbody6}. We therefore adjusted the code to force the time step to always be larger than this value. This adjustment only affects the small number of stars that approach extremely close to the SMBH. Because these stars use time steps which are larger than would naturally be chosen by \textsc{nbody6} we analyzed them to determine whether they were being integrated precisely enough for our purposes. One possible result of imprecise integrations of close approaches is stars that spuriously attain positive energy with respect to the entire system. Stars that have small energy errors on close approaches but that still are representative of a typical star in a galactic center are acceptable for our purposes. Stars that are completely ejected from our system erroneously are unacceptable, because they alter the stellar distribution significantly. However, it is possible for stars to be ejected from our system for natural reasons, including having a close approach with another star while close to the SMBH. The analysis of several hundred stars that were ejected from our simulations found that all of them had a close approach with another star which accounted for the ejection. Additionally, the total energy conservation in our simulations remained within several percent, which is sufficient for our purposes.

\indent \textsc{hnbody} \citep{1999DDA....30.0302R}\footnote{See http://www.hnbody.org} is a symplectic $N$-Body integrator designed specifically for cases in which there is a central massive object that dominates the orbits of all others. However, it also includes adaptive time step integrators, and we chose to use a fourth order Runge-Kutta integrator to better follow extremely high eccentricity orbits. It includes a module that adds post-Newtonian corrections to the Newtonian force calculations of up to order 1PN. This order of correction accounts for general relativistic precession, but does not include gravitational radiation.  However, there is a module in \textsc{hnbody} called \textsc{hndrag} that can also include drag forces.  One of the included options is \textsc{grdrag} \citep{2006ApJ...640..156G}, which models gravitational radiation using the equations of Itoh et al. (2001) as a drag force on the orbit. We use \textsc{hnbody} with the \textsc{grdrag} module to follow the three-body interaction of the BHBs with the SMBH. \\

\subsection{Initial Conditions}

\begin{table*}[t]
\renewcommand{\arraystretch}{3}
\begin{center}
\caption{\label{tab:param}}
\begin{tabular}{|l||c|c|c|c|}
\hline
Simulation & 1k & 2.5k & 5k & 10k\\
\hline \hline
Mass of SMBH (M$_{\odot}$) & $10^{3}$ & $2.5\times10^{3}$ & $5\times10^{3}$ & $10^{4}$\\
\hline
Number of Stars & $307$ & $680$ & $2000$ & $4400$\\
\hline
$d_{\rm init}$ for BHB (AU)  & $500$ & $700$ & $1500$ & $2500$\\
\hline
$d_{\rm max}$ for Stars (AU) & $2000$ & $3000$ & $6000$ & $9000$\\
\hline
Simulation Time (yrs) & $4\times10^{4}$ & $6\times10^{4}$ & $1.12\times10^{5}$ & $3.5\times10^{5}$\\
\hline
\end{tabular}
\end{center}
\label{tab:var}
\tablecomments{\small{Basic parameters that change for each simulation size. The initial semi-major axis of the BHB ($d_{\rm init}$) is set so that the relaxation time and Kozai-Lidov time are equal. The maximum semi-major axis for stars ($d_{\rm max}$) is set to be $\sim 4 \times d_{\rm init}$. The number of stars is set using $d_{\rm max}$ and the number density of stars. The simulation time is set to two relaxation times for the BHB.}}
\end{table*}

Our initial conditions are constrained by the necessity that we be able to simulate these conditions in a reasonable amount of time and with acceptable accuracy. We are therefore limited in the size and complexity of the systems we can simulate. In order to explore the parameter space available to us we have opted to simulate four successively larger systems in an effort to deduce possible trends and make predictions for more realistic conditions. The parameters that vary between these simulations are listed in Table \ref{tab:param}. 

\indent The masses of the SMBHs in our simulations range from $10^{3}$ M$_{\odot}$ to $10^{4}$ M$_{\odot}$. We are constrained not by the number of particles in our simulations, but by the mass ratio between stars and the SMBH. The BHB is placed around this SMBH at a semi-major axis $d_{\rm init}$ such that the KL timescale is equal to the relaxation timescale for the BHB. A BHB further from the SMBH than this will not have time to undergo KL oscillations. Placing the BHB further in is unnecessary as it will migrate inwards due to dynamical friction because the BHB is more massive than the surrounding stars. The maximum semi-major axis for stars is then chosen to be $\sim 4 \times d_{\rm init}$ so that the BHB is deeply imbedded in a distribution of stars. The number of stars is set using the mass interior to the radius of influence along with the number density of stars. Finally, the simulations were run initially for two relaxation times for the BHB particles. We then extended any simulations in which the BHB did not either merge or become tidally separated by the SMBH for the length of time necessary for this to happen.

\indent There are several parameters that remain constant between our simulations. These values were chosen to be representative of typical values, though again, very little is known about black hole binaries. The masses of star particles are set uniformly to $1$ M$_{\odot}$. The number density of stars is a single power law $n \propto r^{-\alpha}$ with $\alpha = 2$ (See Table 1). The BHB particles are each 20 M$_{\odot}$ consisting of two 10 M$_{\odot}$ black holes set at a semi-major axis of 1 AU. 

\indent Each set of initial conditions is created following the parameters laid out in Table 2. For example, the semi-major axes of the stars are drawn randomly from the number-density law $n \propto r^{-2}$ with a maximum at $d_{max}$. The eccentricities for all objects are randomly drawn from a thermal distribution such that the probability of choosing a star with orbital eccentricity between $e$ and $e+de$ is $dP(e) = 2ede$. The true anomalies are drawn randomly from a properly time-weighted distribution, giving more weight to the positions at which the star or BHB remains the longest.  Finally, all other angles including inclination are drawn from an isotropic distribution. The initial interior orbital elements of the BHBs are drawn from the same distributions as the parameters of their COM orbits with the exception of their semi-major axes.  For simplicity, these are fixed at 1 AU.  A thermal distribution for stars is expected for nearly Keplerian orbits \citep{1987gady.book.....B}, while a thermal distribution for binary orbits is motivated by studies of Galactic field binary systems \citep[e.g.,][]{2001ApJ...555..945K}. If these BHBs are mostly formed from evolved post common envelope systems, the eccentricities should be closer to circular.

\section{Results}

For each of the four simulation sizes, we produced ten sets of initial conditions. These initial conditions are integrated forward through at least two relaxation times for the BHB particle. These integrations produce a track for the COM BHB particle. For each of these 40 COM BHB tracks we then produced 10 sets of initial conditions for the orbit of the individual members of the binary. We then integrated these forward, along with the SMBH, while forcing the binary to follow the COM track. The result is 100 iterations of BHB orbits for each simulation size, for a total of 400 different simulations. We will refer to the simulations by 3 numbers. The mass of the SMBH, the COM track number, and the iteration number. Thus for example simulation 5k\_3 would be one based around a $5\times10^{3}$ M$_{\odot}$ SMBH, and it would be the 3rd of 10 COM evolutionary tracks. Simulation 5k\_3\_4 would be the 4th of 10 iterations of the interior BHB orbit of the COM evolutionary track from simulation 5k\_3. These results are then analyzed to determine which binaries have undergone KL oscillations and which have merged due to GW emission. 

\begin{figure*}[htb]
\centering
\includegraphics[height = 5in, width = 7in]{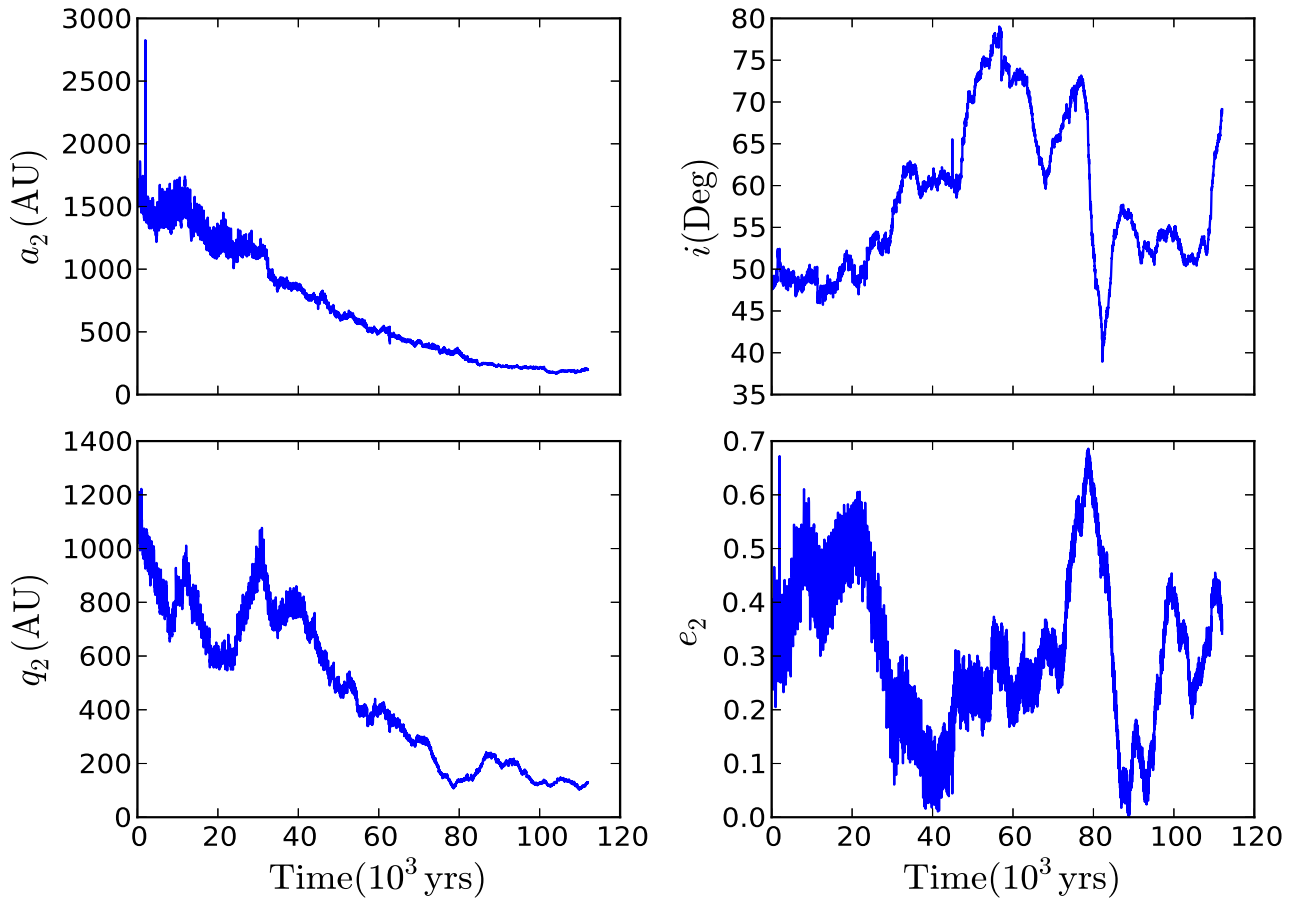}
\caption{ \label{fig:example1} \small{Example output of the COM orbit of the BHB from simulation 5k\_3 around a $5\times10^{3}$M$_{\odot}$ SMBH converted into Keplerian orbital elements.  The superorbit semi-major axis ($a_2$), eccentricity ($e_2$), and pericenter ($q_2$) are plotted along with the inclination of the COM to a reference plane. The effects of the stellar potential on the orbit of the BHB are clearly visible.  The eccentricity and inclination wander significantly over the simulation. The semi-major axis shrinks from 1500 AU to 200 AU due to mass segregation.}}
\end{figure*}

\indent  Figure \ref{fig:example1} shows the results of simulation 5k\_3, a typical evolution of the COM orbit of a BHB around a $5\times10^{3}$ M$_{\odot}$ SMBH. The position and velocity are converted to Keplerian elements. The orbit is not quite Keplerian, but because the SMBH dominates the potential, Keplerian elements describe the orbit reasonably well. The BHB starts at a semi-major axis of 1500 AU and an eccentricity of 0.4. Over the course of 112,000 yrs it exchanges energy and angular momentum with the surrounding stars. The eccentricity and inclination wander significantly. Due to dynamical friction the BHB sinks towards the SMBH, as its mass is much larger than the surrounding stars. By the end of the simulation the semi-major axis is around 200 AU. The mass segregation effect is very pronounced in our simulations in general. Using a more realistic initial mass function (IMF) for the stars in our simulations could slow down this effect, as for \citet{2014ApJ...794..106A}, who showed that in their simulations black holes in the galactic center segregate at a rate similar to the energy relaxation timescale of the dominant stellar population. 

\begin{figure*}[htb]
\centering
\includegraphics[height = 5in, width = 7in]{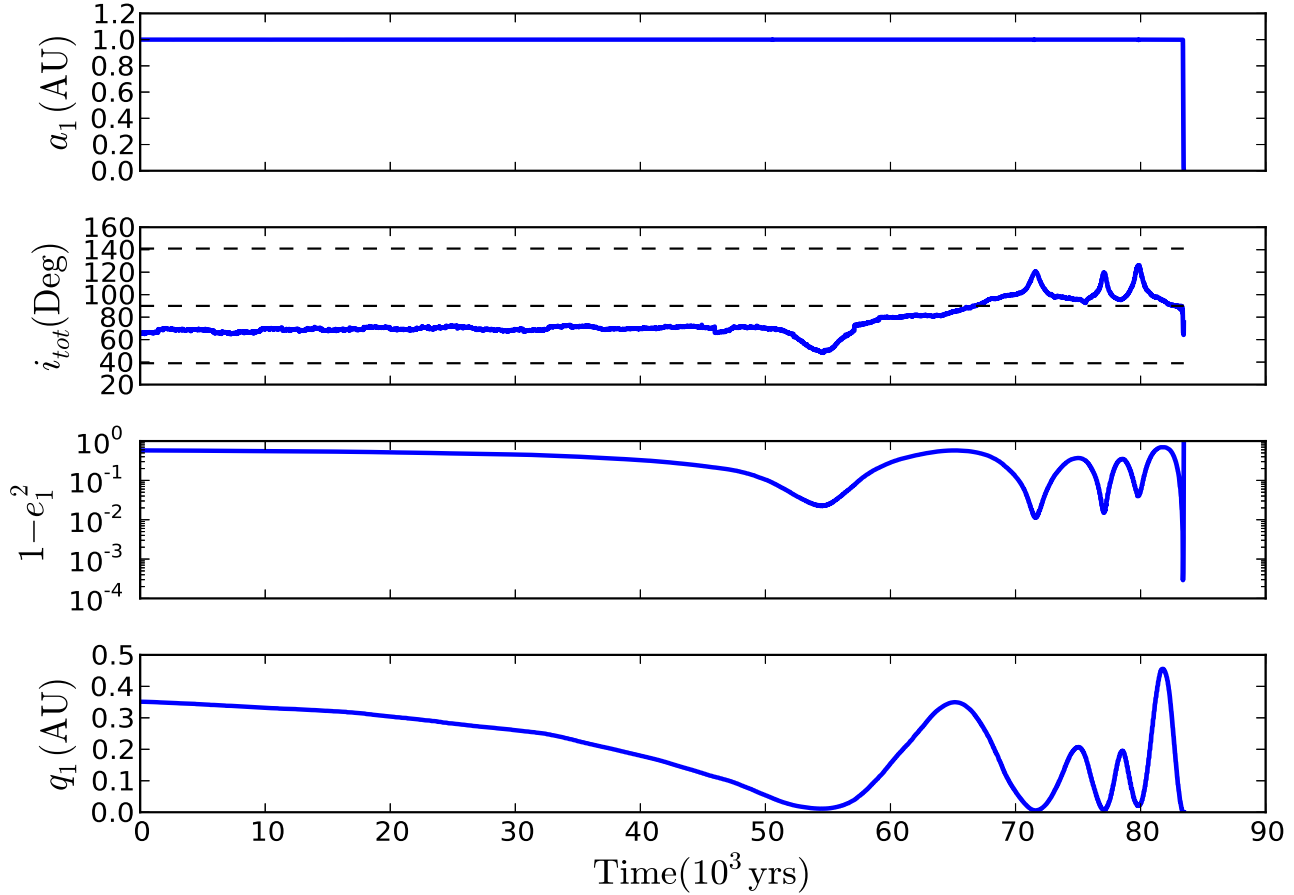}
\caption{ \label{fig:example2} \small{Example of the evolution of the internal orbital elements of a BHB from simulation 5k\_3\_2.  The interior semi-major axis ($a_1$), eccentricity ($e_1$) and pericenter ($q_1$) are plotted along with the total inclination ($i_{tot}$). The eccentricity is plotted as $1-e_{1}^{2}$ to better show the different peaks as well as to relate to $T_{\rm GW}$ and $T_{\rm KL}$. The dashed lines in the plot of $i_{tot}$ show the range of inclinations that lead to KL oscillations in the quadrupole limit, with the critical angle at $90^{\circ}$. KL oscillations are readily apparent in the eccentricity and total inclination. The BHB merges at the end of this simulation as GW emission becomes efficient when the eccentricity reaches a large enough value.}}
\end{figure*}

\indent Figure \ref{fig:example2} shows the results of simulation 5k\_3\_2, an example of the evolution of the internal orbital elements of a BHB. Many of the important features of our simulations can be seen in this example. The total inclination between the interior orbit of the BHB and the exterior orbit around the SMBH grows from about $65^{\circ}$ to just over $90^{\circ}$ and shows both the random walk due to the effects of the stellar potential as well as the oscillations due to the KL mechanism. The overall change in $i_{\rm{tot}}$ maps directly to the inclination of the COM orbit of the BHB. This would likely be damped due to the effects described in Section 2.1. The eccentricity mirrors the KL oscillations of the total inclination. The closer the total inclination approaches to $90^{\circ}$ the more extreme the eccentricity and inclination oscillations become. A little more than $8\times10^{4}$ years through the simulation, the eccentricity reaches a value of 0.99985 and GW emission causes the BHB to inspiral in less than 100 years. Because most of the orbital energy of the BHB is lost near the pericenter passage, the GW emission acts as an impulsive force, driving the BHB to lower eccentricity as it spirals inward. This example also shows the rate of the KL oscillations increasing as the BHB moves closer to the SMBH over time.

\begin{figure*}[htb]
\centering
\includegraphics[height = 4.5in, width = 7in]{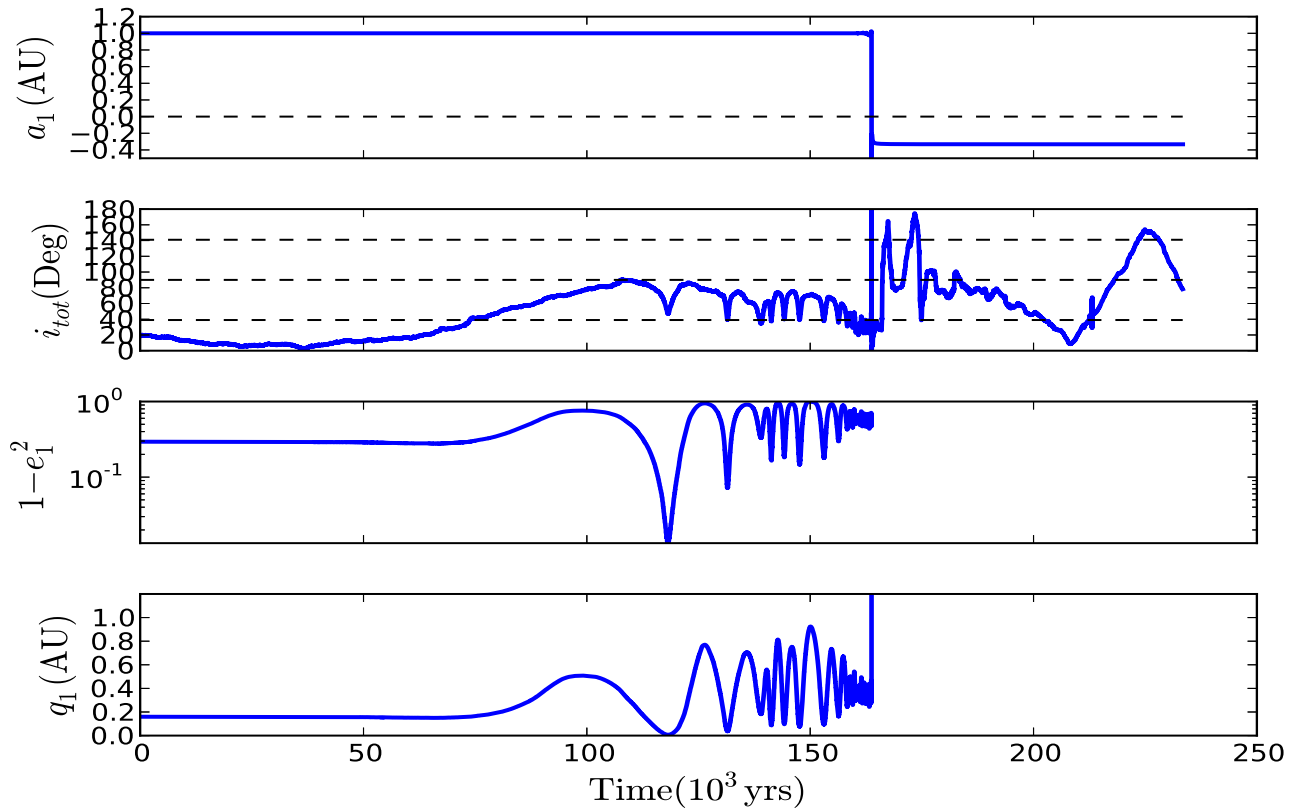}
\caption{ \label{fig:example3} \small{Example of the evolution of the internal elements of a BHB from simulation 10k\_4\_9. The dashed line in the plot of $a_{1}$ is the line at which a binary becomes ionized. This BHB is tidally separated before it can merge due to the KL mechanism. At around $1.6\times10^{5}$ yrs the semi-major axis becomes negative and the eccentricity larger than unity indicating ionization.}}
\end{figure*}

\indent Figure \ref{fig:example3} shows an example of the evolution of the orbital elements of a BHB orbiting a $10^{4}$ M$_{\odot}$ SMBH that is tidally separated before it can merge due to the KL mechanism. The evolution is much the same as that in Figure \ref{fig:example2}, however in this case the BHB never reaches the correct orientation for the KL oscillations to bring it to high enough eccentricity for GW emission to become efficient. Instead, before this can happen, the BHB approaches within 10 AU of the SMBH, causing the BHB to be ripped apart by extreme tidal forces. Inspiral and tidal separation are the two possible outcomes of our simulations. Most simulated BHBs reach one of these outcomes within two relaxation times. Regardless, we follow all BHBs until they meet their final fate.

\begin{figure*}[htb]
\centering
\includegraphics[height = 4.5in, width = 7in]{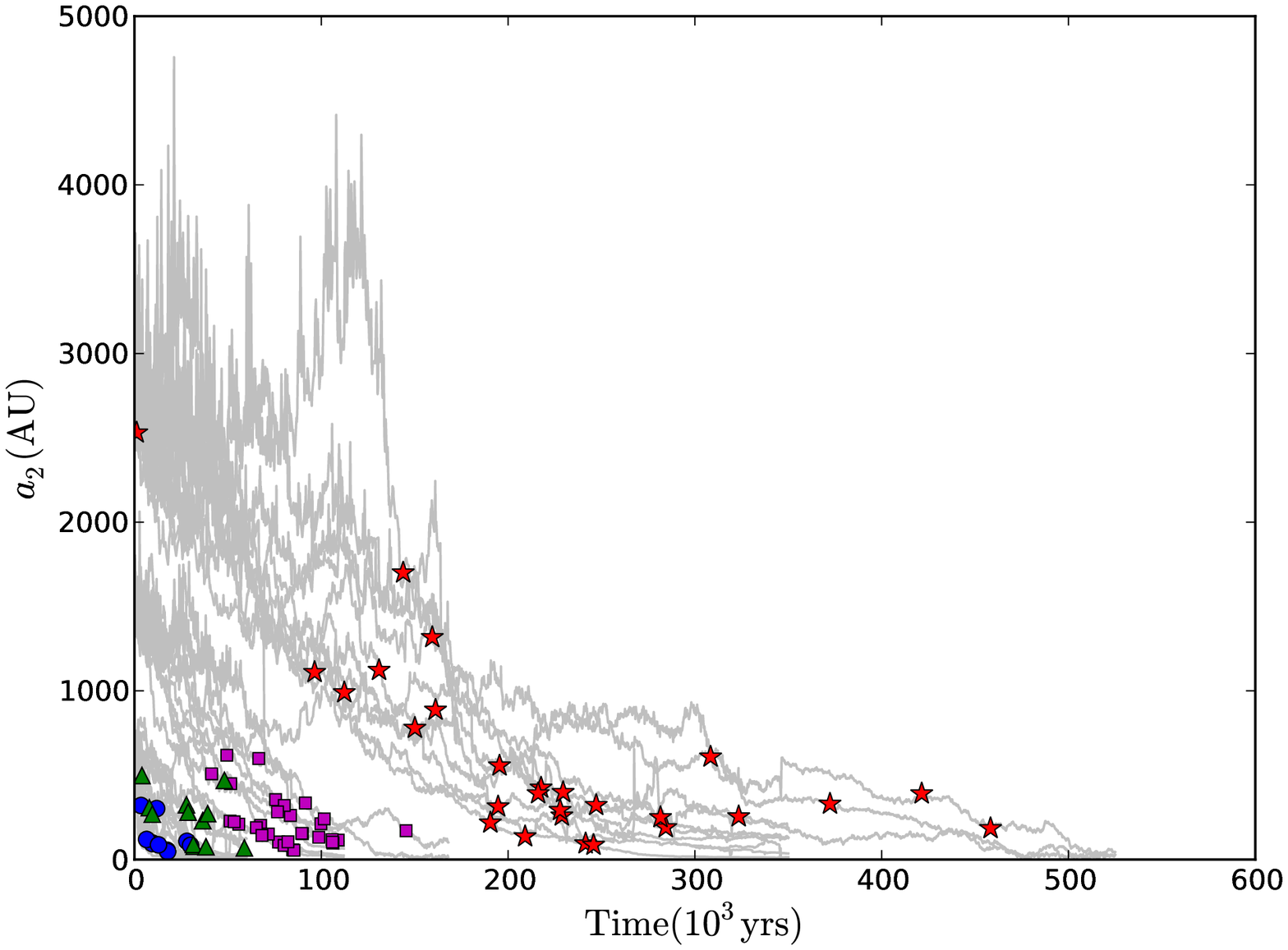}
\caption{ \label{fig:tracks} \small{BHB mergers plotted as a function of the superorbit semi-major axis ($a_{2}$) and time at which they merged. Blue circles are mergers from 1k simulations, green triangles from 2.5k simulations, purple squares from 5k simulations, and red stars from 10k simulations. These are plotted on top of the tracks of the evolution of $a_{2}$ over time. Many of the mergers occur relatively close to the SMBH. There is in general an absence of mergers at early times.}}
\end{figure*}

\begin{figure}[htb]
\includegraphics[width=.50\textwidth,left]{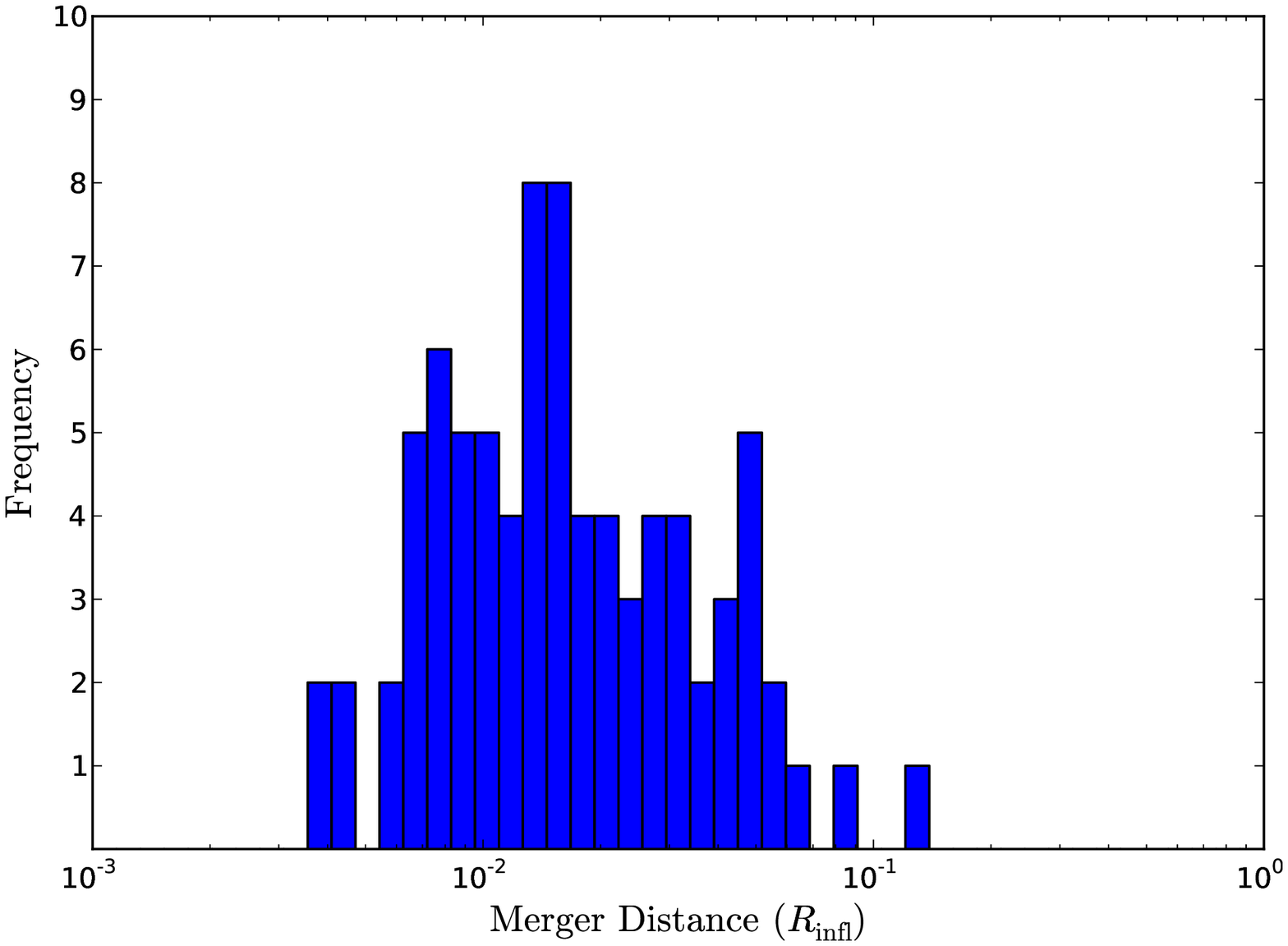}
\caption{ \label{fig:hist} \small{Histogram of the frequency of BHB merger at various distances from the SMBH relative to $R_{\rm infl}$. The most common merger distance is $\sim0.01R_{\rm infl}$.}}
\end{figure}

\indent All of our simulations show BHBs undergoing KL cycles during their evolution. However, not all of our simulations end in merger. Figure \ref{fig:tracks} shows the semi-major axis and time at which each BHB merged superimposed on the evolution of the superorbit semi-major axis over time. Many of the mergers occur when the BHB is relatively close to the SMBH. There is in general an absence of mergers at early times and large semi-major axis. There may also be fewer mergers at very late times and small semi-major axis (See Figure \ref{fig:hist}). The absence of mergers at early times may be due to our initial conditions. If there is a full loss cone \citep{2006astro.ph..1520H,2015ApJ...804..128M} around the SMBH, BHBs from outside our simulation region would enter this region while already undergoing KL oscillations and possibly add to the number of mergers in the outer region. This outcome depends heavily on the details of the mass segregation process \citep[e.g.,][]{2006ApJ...645L.133H,2009ApJ...697.1861A,2014ApJ...794..106A}. The possible absence of mergers at late times is most likely due to the number of BHBs that are tidally separated. Relatively fast mass segregation brings our BHBs close enough to be separated before they have time to merge. With a slower mass segregation rate BHBs would linger longer and the number of mergers would likely increase.

\begin{figure*}[htb]
\centering
\includegraphics[height = 4.5in, width = 7in]{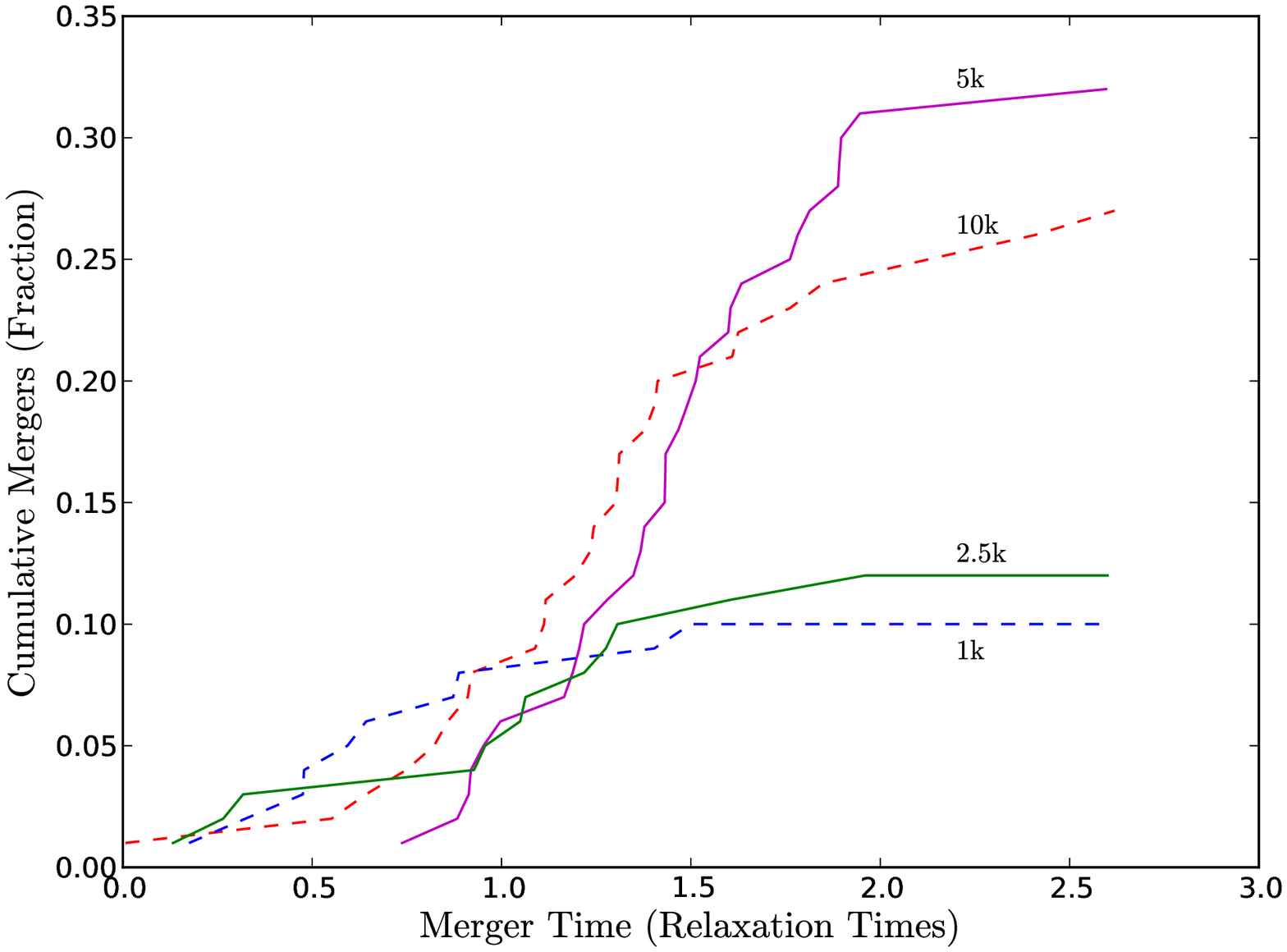}
\caption{ \label{fig:cumulative} \small{Fraction of simulations of each of the four classes that have merged over time. The time is scaled to the relaxation time for each simulation class (See Table 2). The simulation classes are marked on the plot. A significant fraction of the simulations of all classes merge over several relaxation times. }}
\end{figure*}

\indent Figure \ref{fig:cumulative} shows the fraction of BHBs that merge over time for each of the four classes of simulation. For each of the simulation sizes, a significant fraction of the BHBs merge. For $10^{3}$ M$_{\odot}$ SMBHs, 10\% of the BHBs merge, in the 2.5k simulations 12\% merge, in the 5k simulations 32\% merge, and in the 10k simulations 27\% merge. The merger rate appears to be fairly steady. As stated previously, it is possible that the rate may be slightly lower near the beginning and end of the simulations than in the middle. The merger rate is clearly lower for the 1k and 2.5k simulations, while it may be slightly higher for the 5k simulations than for the 10k simulations. 

\indent There could be several reasons for the merger rate to be lower in the smaller simulations. One in particular can be seen from the different dependencies of several of the timescales (Table 1). For a smaller SMBH, $T_{\rm KL}$ gets longer, while $T_{\rm ER}$ gets shorter. This allows fewer KL oscillations per relaxation time, meaning fewer chances to reach high eccentricity. The similar merger rates for the 5k and 10k simulations suggests that these simulations may be high enough resolution to capture the main elements important to mergers from this channel. 

\indent The change in $i_{\rm{tot}}$ in all of our simulations is likely larger than is realistic, as discussed in Section 2.1. Our conclusions as to the number of BHBs that would reach the critical inclination required to merge should therefore be considered an upper limit. However, the general behavior of these BHBs in this dense stellar potential is still instructive. The evolution of the COM orbits of the BHBs are still followed correctly, and this allows an investigation of the interplay of all of the various timescales relevant to this scenario in a simulation for the first time.

\subsection{Detection Rate}
Using our results to find an accurate detection rate per volume from this channel is not possible at this time. The fraction of BHBs that merge in our simulations is likely an upper limit as previously discussed. Additionally, the systems we were able to simulate are much smaller than realistic systems, and we do not yet have enough data to say with confidence that we see a strong trend with increasing M$_{\rm SMBH}$. However, it is still instructive to look at how our results might translate to a larger system. The following is therefore a simple upper limit estimate meant to guide understanding of the possible importance of this channel in BHB merger rates.

\indent The maximum rate of mergers in our simulations is about $15\%$ per relaxation time for the 5k simulations. The scaled relaxation time for a Milky Way type galaxy with a SMBH of mass $\sim 4\times10^{6}$ M$_{\odot}$ can be found by equating $T_{\rm KL}$ and $T_{\rm ER}$ from Table 1 as we have to set the initial conditions in our simulations. This relaxation time is $\sim 6\times10^{8}$ yrs at about 1.2 parsecs. Out to this distance, which is well inside $R_{\rm infl}$, we could expect to see a total mass of stars of $\sim 2\times10^{6}$ M$_{\odot}$ \citep{2009JKAS...42...17O}. For an average stellar mass of 1 M$_{\odot}$ that would be $\sim 2\times10^{6}$ stars. We then need to know the fraction of those stars expected to be black holes, $f_{\rm BH}$, as well as the fraction of those black holes expected to be in binaries, $f_{\rm bin}$. Estimates of $f_{\rm BH}$ are very uncertain and range from 0.001 \citep{2006ApJ...645.1152H,2014ApJ...794..106A} to 0.016 \citep{2000ApJ...545..847M}. The former estimate is based on population synthesis models, while the latter takes into account a strong mass segregation effect, increasing the number of black holes toward the center of the galaxy. Estimates of $f_{\rm bin}$ are also highly uncertain, but one estimate from \citet{2004ApJ...611.1068B} found that the binary fraction should be $\sim 20\%$, also from population synthesis models. From these values we estimate the total merger rate:

\begin{multline}
\frac{\Delta N_{\rm merge}}{\Delta t} = \frac{f_{\rm merge}N_{\rm stars}f_{\rm BH}f_{\rm bin}}{\rm Merger\, Time} = \\\frac{0.15\times2\times10^{6}\times0.016\times0.2}{6\times10^{8} \mbox{yrs}} \approx 2\, \rm{per\, Myr}.
\end{multline}

\noindent Here $f_{\rm merge}$ is the fraction of BHBs that merged, $N_{\rm stars}$ is the number of stars expected within the volume and the merger time is the relaxation time. 

\indent This estimation results in a rate $R \approx 2$ per Myr per Milky Way Equivalent Galaxy (MWEG). In order to convert this rate to a detection rate per volume we use an extrapolated density of MWEG's of 0.0116 Mpc$^{-3}$ \citep{2008ApJ...675.1459K}. This gives us a detection rate of $N \sim 100$~Gpc$^{-3}$~yr$^{-1}$. This rate lands in the middle of the range of expected values of the total rate of stellar mass black hole mergers discussed in Section 1. This is  a very rough upper limit estimate, as we have used optimistic values for the merger fraction and the black hole fraction. This rate also depends on a steady supply of new black holes entering the galactic center, which is uncertain. Finally, it should be taken into account that this is but one route to merger, which should be considered in addition to all others.  

\subsection{Eccentric Mergers}
One potentially interesting consequence of a BHB merger due to strong KL oscillations is the possibility of a merger that enters the LIGO frequency range while still being somewhat eccentric. Such mergers would likely have significantly different waveforms and therefore require different templates and search algorithms to find \citep[e.g][]{2013PhRvD..87d3004E,2013PhRvD..87l7501H}. There have therefore been various attempts to determine how common these eccentric mergers might be. The frequency typically considered is 10 Hz because this is near the lowest frequency that current ground-based detectors can achieve. Mergers are commonly considered 'eccentric' in this scenario when they reach this frequency with an eccentricity $\gtrsim 0.1$. In scenarios involving field binaries \citep[e.g.,][]{2008ApJ...676.1162S,2015ApJ...814...58D}, mergers will be very close to circular by the time they reach the LIGO frequency range. Mergers involving few-body encounters are more complex. Studies find that these should typically result in circular mergers \citep[e.g.,][]{2006ApJ...640..156G,2014MNRAS.441.3703Z}, though some fraction could still retain significant eccentricity  \citep{2014ApJ...784...71S}. Dynamic scenarios involving two-body capture \citep{2009MNRAS.395.2127O} or resulting in hierarchical triple systems that undergo KL oscillations \citep{2002ApJ...576..894M,2003ApJ...598..419W,2012ApJ...757...27A,2014ApJ...784...71S,2016ApJ...816...65A} are commonly found to have non-negligible chances of eccentric merger. Estimates of up to $\sim30\%$ of mergers being eccentric have been predicted \citep{2003ApJ...598..419W}. 

\indent In order to determine whether any of our simulations would result in an eccentric merger we use a simple prescription. We evaluate the eccentricity at 10 Hz. In order for our BHBs to orbit on that time scale they need to reach a semi-major axis of $\approx 10^{-6}$ AU. Because the energy loss in these initially highly eccentric systems is nearly impulsive, reaching that semi-major axis while still remaining eccentric requires reaching an even smaller pericenter as the merger begins. Of the 81 mergers across all 400 of our simulations, only one reaches a pericenter closer than $10^{-6}$ AU: simulation 5k\_1\_9 reaches a pericenter of $4.5$x$10^{-7}$ AU, which results in an $e=0.58$ at 10 Hz. From this relatively small sample we conclude that eccentric mergers due to KL oscillations observable by LIGO are indeed possible, though likely to be rare.

\section{Other Considerations}

The merger of BHBs due to KL oscillations in galactic centers is a complex process subject to many uncertainties. Our work here is a step forward in understanding the most important processes responsible for this outcome. However, there are still several opportunities for improving upon this work. Here we examine those processes that may positively or negatively impact the merger rate of BHBs and which may be incompletely simulated in our work.

\indent One of the most important effects that we did not follow accurately in our simulations is the damping of the change in mutual inclination of the BHBs to the SMBH due to the shorter timescale of the KL oscillations relative to the timescale of the change in inclination of the superorbit of the BHBs. We discussed this effect in Section 2.1 and made some progress towards understanding how important this damping might be. However, there is still a great deal of work to be done. Primarily, it will be important to know how much of the inclination change is due to slow, smooth processes such as VRR compared to fast processes such as kicks from close approaches. If kicks from close approaches are an important source of inclination change, it may still be possible to get a moderate change in the mutual inclination. Even if this is not the case, there is likely to be some range of initial inclinations that can reach the critical inclination for merger. Determining this range will add to our understanding of the fraction of BHBs that are able to undergo merger due to KL oscillations significantly. A range of even a few degrees would increase the fraction of mergers significantly from a static inclination scenario. We will undertake a more thorough investigation of these issues in a later paper.

\indent Another set of important possible issues arise due to the relatively small number of particles and low mass of the SMBHs in our simulations compared to MWEG nuclei. One such issue relates to the varying dependencies on $M_{\rm SMBH}$ of important timescales (See Table 1). Most clearly, $T_{\rm KL}$ decreases with $M_{\rm SMBH}$ while $T_{\rm ER}$ and $T_{\rm VRR}$ increase with $M_{\rm SMBH}$. We therefore expect that BHBs around a more massive SMBH would experience more KL oscillations at a particular distance and inclination. This would tend to increase the likelihood of merger in larger systems. This particular effect may be the reason for the varying merger rates in our 1k and 2.5k simulations compared to our 5k and 10k simulations.

\indent A similar issue, but with possibly a negative effect on the merger rate, is the relative importance of binary ionization in a larger system. In our simulations the BHBs are hard at almost all distances from the SMBH. In the 10k simulations, a typical star would have greater kinetic energy than the internal binding energy of our BHBs only within 100 AU of the SMBH. However, this distance increases in proportion to $M_{\rm SMBH}$ inside $R_{\rm infl}$ while $R_{\rm infl}$ only increases as the square root of $M_{\rm SMBH}$ (See Eq. 3). Therefore we would expect a larger proportion of BHBs to be soft in larger systems, increasing the importance of this mechanism.

\indent The relatively small numbers of stars in our simulations leaves open the possibility of stochastic processes playing a part in our results. \citet{2005LRR.....8....8M} describes various scenarios where small particle numbers do not adequately follow physical processes such as loss cone refilling and binary hardening. Large particle number experiments may be necessary in order to confirm the behavior seen in our simulations.

\indent The separation of our simulations into two parts is also a simplification worthy of scrutiny. Because of this simplification, close approaches of stars to a BHB have no effect on the internal properties of the BHB. If these close approaches were to be taken into account properly the effect would on average be to harden BHBs which are hard, soften BHBs that are soft \citep{1977ComAp...7...43H}, or to have a strong three-body interaction possibly scramble the internal properties of the BHB if the interaction is extremely close. Because our BHBs are almost always hard, we are likely missing a net hardening effect, which could increase the merger rate slightly. A series of close approaches could also add an additional source of mutual inclination variation, which would not be affected by the damping discussed in Section 2.1. We found very close approaches to be quite rare (See Section 3.1), but it is possible that they could still have played a role.

\indent One additional effect that is due to the splitting of our simulations is that each BHB is repositioned slightly at the end of each \textsc{nbody6} output step. This results in small jumps in the orientation of the BHB instead of a smooth transition. The orientation changes are quite small and happen approximately every 10 simulated years. This effect is unavoidable using this simulation method. Because it happens on a timescale much shorter than $T_{\rm KL}$ the repositioning should not be a large effect.

\indent Our simulations also made a few simplifying assumptions about conditions in a galactic center that could have an effect on the merger rate. We used a simple IMF consisting entirely of 1 M$_{\odot}$ stars, a single power law number density for stars with $\alpha = 2$, an equal mass BHB, and a set BHB separation of 1 AU. Using a more realistic IMF that includes massive compact objects as in \citet{2006ApJ...645L.133H} or \citet{2014ApJ...794..106A} could slow mass segregation. A shallower number density law for objects in the galactic center as observed in for example \citet{2010ApJ...708..834B} would increase $T_{\rm ER}$ though the actual density distribution in the galactic center is still debated and quite complex \citep[e.g.,][]{2005PhR...419...65A,2010ApJ...718..739M}. 

\indent Recent population synthesis models find BHBs to on average have components of fairly equal masses \citep{2015arXiv151004615B}. Unequal mass BHBs introduce octupole order terms into the KL mechanism, which could slightly alter merger times \citep[e.g.,][]{2011ApJ...742...94L}. The distribution of semi-major axes of BHBs is very uncertain, though \citet{2002ApJ...572..407B} found an average of around 10 R$_{\odot}$ and a study of solar neighborhood binaries found a log-normal distribution of periods \citep{1991A&A...248..485D}. Our choice of an initial separation of 1 AU is if anything a conservative estimate. Choosing a smaller initial separation would reduce the GW merger time considerably. However, this could also reduce the effect of encounters on inclination changes, thus leading to a smaller number of mergers.

\indent One of the most uncertain variables when determining the merger rate of BHBs is the number of BHBs expected to be in the galactic center, as well as how quickly they are repopulated. We do not directly simulate this, instead taking an optimistic value from the literature. Processes that can replenish the BHB population in the galactic center include diffusional processes \citep[e.g.,][]{2000ApJ...545..847M,2009ApJ...700.1933H}, disruption of triple systems \citep[e.g.,][]{2009ApJ...698.1330P}, in-situ star formation \citep[e.g.,][]{2008ApJ...674..927A}, and three- or four-body interactions \citep[e.g.,][]{2002ApJ...576..894M}.

\indent Several recent observational studies have found a surprising overabundance of low mass X-ray binaries \citep{2005ApJ...622L.113M} and excess in hard-X-ray emission \citep{2015Natur.520..646P} in the galactic center. Both of these studies conclude that this overabundance challenges our understanding of binary formation and evolution in the galactic center. One possible explanation for this overabundance of accreting massive objects is an excess of highly eccentric binaries that leads to mass transfer. It is therefore possible that a process similar to that discussed in this work is responsible for these observations \citep{2015ApJ...799..118P}. The KL mechanism could just as easily bring a binary consisting of a main sequence star and compact object to high eccentricity as a BHB. Though there are differences, including stellar evolution and tidal friction that are present in these systems, the KL mechanism remains a promising avenue to explore as an explanation for the overabundance of X-ray binaries in the galactic center.

\section{Conclusions}

In this work we examined the effect of the KL mechanism on the merger of BHBs in galactic centers. We used direct $N$-body simulations in order to capture the important processes that play a role in this scenario. Using a unique combination of two $N$-body codes we were able to simulate close approaches of stars to the SMBH as well as the internal orbit of the BHB. We have shown that the KL mechanism plays an important role in the evolution of the orbits of BHBs in galactic centers. Additionally, we found that the merger rate of BHBs is enhanced compared to field binaries by the influence of the SMBH. This rate may be overestimating the effect of inclination changes of the superorbit on the mutual inclination of the BHBs. We have also shown that eccentric mergers in the LIGO frequency band are possible but not common in our simulations. Finally, we discussed possible improvements on this work and suggested that the KL mechanism may be important in explaining the overabundance of low mass X-ray binaries in the galactic center.\\

\indent This paper was supported in part by NASA ATP grant NNX12AG29G. We thank Sverre J. Aarseth and Simon Karl, both of whom were instrumental in enabling us to use \textsc{nbody6} for this project. Conversations with both of them proved extremely valuable. We also thank Fabio Antonini, who brought to our attention the problems with considering our BHBs to be point particles while evolving their COM orbits. His explanation of this subtle problem and support for this paper were very much appreciated. Thanks also to Hagai Perets, Nick Stone, Johan Samsing, and Imre Bartos for their feedback. JV acknowledges funding support from the Graduate School Deans Fellowship and Merit Fellowship from the University of Maryland.

\end{document}